\documentclass[useAMS,usenatbib]{mn2e}
\usepackage{graphicx}
\usepackage{amsmath}
\usepackage{amsfonts}
\usepackage{amssymb}

\title[LMC clusters at intermediate masses]{Probing the LMC age gap at intermediate cluster masses}

\author[E. Balbinot et al.]{E. Balbinot${^1}$\thanks{balbinot@if.ufrgs.br}, B.X. Santiago$^{1}$, L. O. Kerber$^{2}$, 
B. Barbuy$^{2}$ and B. M. S. Dias$^{2}$ \\
$^1$Departamento de Astronomia, Universidade Federal do Rio Grande do Sul, Porto Alegre, RS, Brazil \\
$^2$IAG, Universidade de S\~{a}o Paulo, S\~{a}o Paulo, SP, Brazil}

\begin{document}

\pagerange{\pageref{firstpage}--\pageref{lastpage}}

\maketitle

\label{firstpage}

   \begin{abstract}

   The LMC has a rich star cluster system spanning a wide range of ages
   and masses. One striking feature of the LMC cluster system is the 
   existence of an age gap between 3-10 Gyrs. But this feature is not as 
   clearly seen among field stars. Three LMC fields containing 
   relatively poor and sparse clusters whose integrated colours are 
   consistent with those 
   of intermediate age simple stellar populations have been imaged in BVI with 
   the Optical Imager (SOI) at the Southern Telescope for Astrophysical 
   Research (SOAR). A total of 6 clusters, 5 of them 
   with estimated initial masses $M < 10^4 M_{\odot}$, were studied in these 
   fields. Photometry was performed and Colour-Magnitude Diagrams (CMD) 
   were built using standard point spread function fitting methods. 
   The faintest stars measured reach $V \sim 23$. The CMD was cleaned from 
   field contamination by making use of the three-dimensional 
   colour and magnitude space available in order to select stars in 
   excess relative to the field. A statistical CMD comparison method was
   developed for this purpose.
   The subtraction method has proven to be successful, yielding cleaned CMDs
   consistent with a simple stellar population. The intermediate age 
   candidates were found to be the oldest in our sample, with ages between 
   1-2 Gyrs. The remaining clusters found in the SOAR/SOI have ages ranging 
   from 100 to 200 Myrs. Our analysis has conclusively shown that none of the 
   relatively low-mass clusters studied by us belongs to the LMC age-gap.
   \end{abstract}


\section{Introduction}


The Large and Small Magellanic Clouds (LMC and SMC, respectively) make up a
very nearby system of low-mass, gas-rich and interacting galaxies.
At their distances, both Clouds can be resolved into stars, allowing their
stellar populations and star formation history (SFH) to be studied in
detail. These studies open up the possibility to identify epochs
of enhanced or reduced star formation, and to associate them to the
system dynamics \citep{holtzman,javiel}. 


Star clusters provide an alternative tool to reconstruct the SFH of a galaxy.
The Magellanic Clouds have a large cluster system,
spanning a wide range of properties, such as masses, ages and metallicities,
which, given their proximity, can be determined using different tools
\citep{santiagoiau}.
Although they make up only a small fraction of the stellar mass in a given
galaxy, star clusters have an advantage over field stars in that
they may be modelled as simple stellar populations (SSPs),
facilitating derivation of their main properties. 
Sizeable samples of clusters with available integrated magnitudes and colours 
currently exist. In the LMC, they have been used in association with 
evolutionary SSPs models to derive age and mass distributions and to
reconstruct the Initial Cluster Mass Function (ICMF) and the Cluster Formation
Rate (CFR) \citep{hunter, grijs, parm}. 

One striking and undisputable feature of the LMC cluster system
is the so-called age gap. 
It was proposed by \citet{jensen}, as a lack of clusters in the
range $4 \leq \tau \leq 10$ Gyrs. Several candidates to fill this gap were
proposed and discarded since then \citep{saraj,rich}.
Evidence for the gap is present in the CFR
reconstruction by \citet{parm}. An age-gap is also tentatively seen in some
reconstructions of the LMC SFH based on field stars, although not
systematically in all fields studied, and not
as clearly as in the cluster system \citep{javiel,noel}.
The lack of a clear age gap among field stars suggests that it may be 
less pronounced among lower mass clusters 
($M < 10^4 M_{\odot}$) as well, which tend to 
be systematically unfavoured in current magnitude-limited cluster samples 
in the LMC. Thus, the sample of candidate clusters with well determined 
ages must be pushed towards fainter limits than has been previously done.

The most reliable cluster age determinations require CMD analysis.
Confrontation of observed CMDs with theoretical ones, built from
stellar evolutionary models, provide the best tool for age determination
of single clusters, either with HST or from the ground 
\citep{baume,piatti09,piatti07}. Building a cluster CMD, however, is a more
costly task than obtaining integrated photometry, as it requires
larger telescopes with high pixel sampling and good seeing conditions.
Therefore, selection of candidates to fill the LMC age gap should be based
on the afore mentioned samples with integrated photometry.
Furthermore, poor and sparse clusters suffer from more severe field
star contamination on their CMDs, something that leads to a
bias towards rich LMC clusters having detailed CMD analysis available.

In this paper we analyse the CMDs of 6 LMC clusters for which no CMD is 
yet available. Two of these clusters have integrated colours consistent with 
an intermediate age, according to the photometry from \citet{hunter}. 
Five of them are sparse and unconspicuous when compared to previous samples. 
Our main goal is to probe relatively poor and intermediate-age clusters 
in search of possible examples that may have ages within the age gap. They 
have been imaged with the Optical Imager (SOI) at the Southern Telescope for 
Astrophysical Research (SOAR), under $\leq 1\arcsec$ seeing and
with excellent PSF sampling. In Sect. 2 we describe the observations
and photometry. In Sect. 3 we present our analysis tools, including a
 method to efficiently decontaminate the cluster CMD from field stars.
This new decontamination method uses the entire information on the 
magnitude and colour space to subtract field stars from the CMD. We 
show the results for each cluster individually. In Sect. 4
we present our final discussion and conclusions.

\section{Data}

\begin{figure}
\centering
 \includegraphics[width=0.4\textwidth]{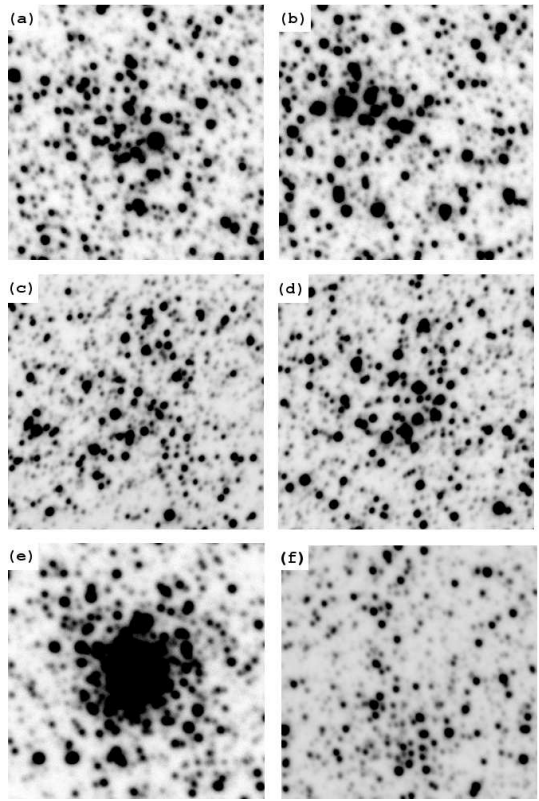}
 \caption{Sections of the final V band images where the clusters are located. The size of each section is $50\arcsec\times50\arcsec$. 
 The cluster in each panel is as follows:
  (a) KMK88-38 (b) KMK88-39 (c) OGLE-LMC0531 (d) OGLE-LMC0523 (e) NGC 1878 (f) OGLE-LMC0214 .In 
  all figures North is up and East is to the right.}
 \label{campos}
\end{figure}

The observations were made in service mode and 
took place in the nights of November 10th 
and December 16th 2007, using the SOAR Optical Imager (SOI) in B, V and I 
filters. 
Fields were imaged around 3 LMC age-gap candidates listed by \citet{hunter}
\footnote{The names OGLE-LMC, KMK88 e BSDL refer to \citet{ogle}, \citet{kmk}, and \citet{bica99}, respectively.} 
:
OGLE-LMC0531, KMK88-38, and OGLE-LMC0169. Another field was observed around
the richer and yet unstudied cluster NGC 1878.
The field around OGLE-LMC0169 was observed under unstable and 
mostly bad seeing for our purposes, which led us to discard the data.
For the remaining fields, the mean seeing was around 0.9 arcsec 
during both nights.

The SOAR Optical Imager (SOI) uses two 2050 $\times$ 4100 px CCDs, 
covering a 5.26 arcmin square field of view (FOV) at a scale of 0.077"/pixel. 
The images have a gap of 10.8" between the two CCDs.
A slow readout was adopted in order to minimize detector noise. A 2$\times$2 
binning was used, yielding a spatial scale of 0.154"/pixel. 

In Table \ref{table:1} we summarize the observation 
log, where we show: target name, filter, seeing, number of exposures, 
and exposure time.
Multiple exposures were taken in order to increase
the signal to noise ratio and reject cosmic rays in the final combined
image. For each cluster, a set of short exposures was also taken in order to
avoid saturation of bright stars. The seeing is the mean full width at half 
maximum (FWHM) of bright stars from the individual exposures. These exposures 
were flatfielded, bias subtracted, mosaiced and combined.

The SOAR/SOI fields included 3 additional star clusters besides 
those taken from \citet{hunter}. They are OGLE-LMC0214, 
OGLE-LMC0523, and KMK88-39. The clusters names are taken from \citet{hunter}.
Fig. \ref{campos} shows the sections (50\arcsec$\times$50\arcsec) of the 
final combined V band images where the clusters are located. Five of them are 
low density and relatively sparse open clusters; 
only NGC 1878 is a richer and more compact object. To our knowledge,
none of them have published CMD data in the literature.

For photometric calibration, two standard fields from \citet{sharpee},
located at the north-eastern arm of the SMC, were observed on each night. 
The observations have distinct air masses ranging from 1.34 to 1.76. 
The magnitudes obtained from these images were used to fit a calibration 
equation for each passband (B,V, and I). We took care to include standards covering
a similar airmass range as the clusters, and under seeing $\le 1\arcsec$. 
The fit residuals were $\sim 0.03 ~ mags$, indicating that both nights were photometric 
for most of the time.

\begin{table}
\caption{Log of observations.} 
\label{table:1} 
\centering
\begin{tabular}{cccc}
\hline
Target & Filter & Seeing(") & Exp. time(s) \\  
\hline
\hline 
OGLE-LMC0531 & B & 0.98 &$ 3\times600 $ \\
	"        & V & 0.85 &$ 3\times200 $ \\
	"        & I & 0.81 &$ 3\times210 $ \\
	"        & B & 0.91 &$ 2\times20  $ \\
	"        & V & 0.86 &$ 2\times15  $ \\
	"        & I & 0.77 &$ 2\times10  $ \\
KMK88-38     & B & 1.01 &$ 3\times600 $ \\
	"        & V & 1.01 &$ 2\times200 $ \\
	"        & V & 1.12 &$ 2\times200 $ \\
	"        & I & 0.91 &$ 3\times210 $ \\
	"        & B & 0.86 &$ 2\times20  $ \\
	"        & V & 0.86 &$ 2\times15  $ \\
	"        & I & 0.86 &$ 2\times10  $ \\
NGC 1878     & B & 1.31 &$ 3\times600 $ \\
	"        & V & 1.15 &$ 3\times200 $ \\
	"        & I & 0.85 &$ 3\times210 $ \\
	"        & B & 1.14 &$ 2\times20  $ \\
	"        & V & 1.05 &$ 2\times15  $ \\
	"        & I & 0.91 &$ 2\times10  $ \\	
\hline 
\end{tabular}
\end{table}

\begin{figure}
\centering
 \includegraphics[width=0.5\textwidth]{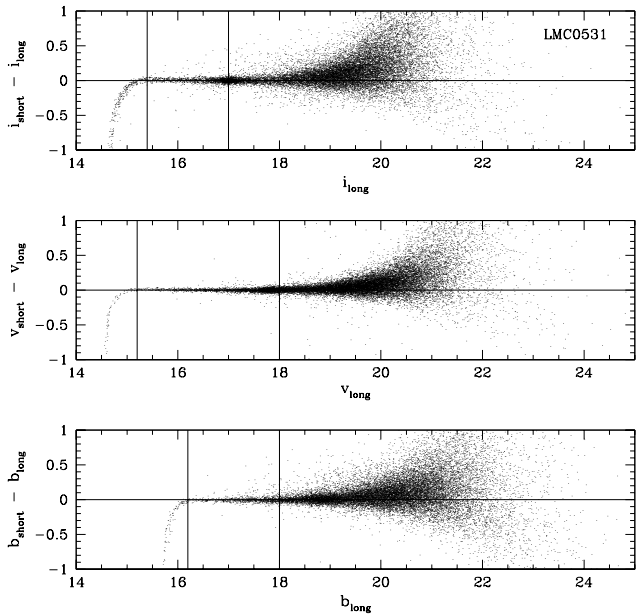}
 \caption{The residual diagrams for OGLE-LMC0531 in the B (lower panel), 
 V(middle panel), and I(upper panel), where we show the difference 
 between the instrumental magnitudes of the \textit{long} and \textit{short} 
 images as a function of the \textit{long} magnitude.}
 \label{res}
\end{figure}

Here we adopt a point spread function (PSF) fit
photometry, the standard approach when dealing with 
crowded fields. The PSF model was computed using the images
from the standard fields in order to avoid crowding effects on the PSF modeling. 
The photometry was carried out using DAOPHOT code from \citet{stetson}. 
The photometry process took place as following: (i) For each cluster we chose the 
deepest and cleaner field for reference, usually the 
long exposure image on the I filter. Automatic detection of sources was carried out 
on this reference field using DAOFIND task, generating a master list of stars. (ii) This master 
list was then used as the position reference. (iii) The photometry on the remaining
passbands then uses this master list, with only small positional offsets applied.

In Fig. \ref{res} we show the residuals between the photometry
of long and short exposure (hereafter \textit{long} and \textit{short}, respectively) 
for the cluster OGLE-LMC0531 as an example. It is clear that our photometry is accurate, 
since bright stars lie very near the zero residual line.
Using this diagram we can determine the \textit{long} saturation 
limit, where the residuals become negative. We can also determine 
the faint magnitude limit where the \textit{short} signal to noise ratio (S/N) 
becomes small and the scatter in the residuals becomes larger. 
A systematic photometric offset of a few hundredths of magnitude where found between 
\textit{short} and \textit{long} and were interpreted as due to small seeing variations. 
They were dealt with by means of aperture corrections applied on the \textit{short} magnitudes.

After photometric homogenization, we merge both \textit{long} and \textit{short} photometry tables. 
In essence, the merger process keeps only \textit{short} stars brighter than 
the \textit{long} saturation limit and only \textit{long} stars fainter than 
the \textit{short} low S/N limit. In the region between the quoted limits we choose
the star with the best magnitude measurement, i.e. the star with the
smaller photometric error.

\begin{figure*}
\centering
 \includegraphics[width=0.4\textwidth]{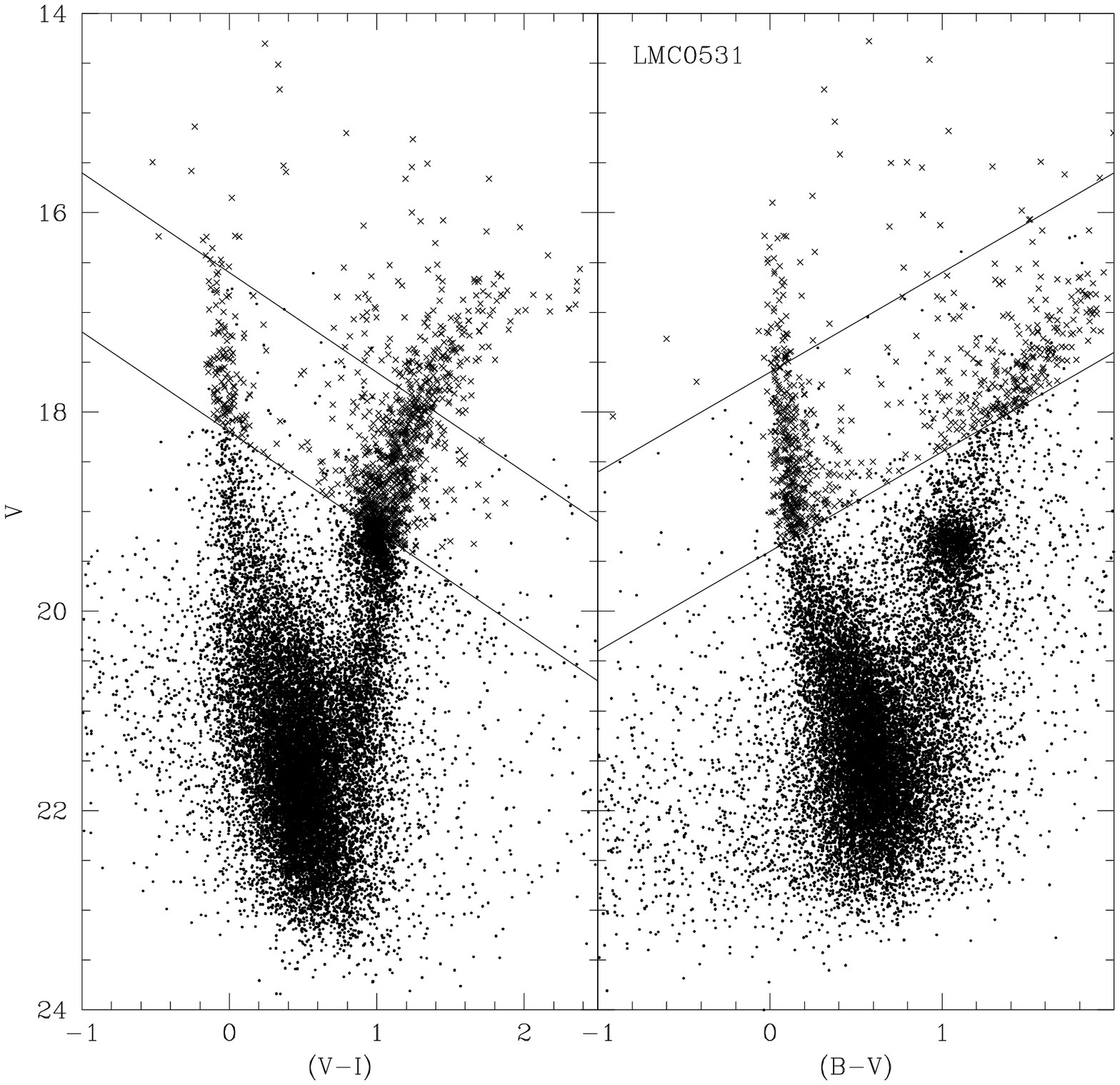}
 \includegraphics[width=0.4\textwidth]{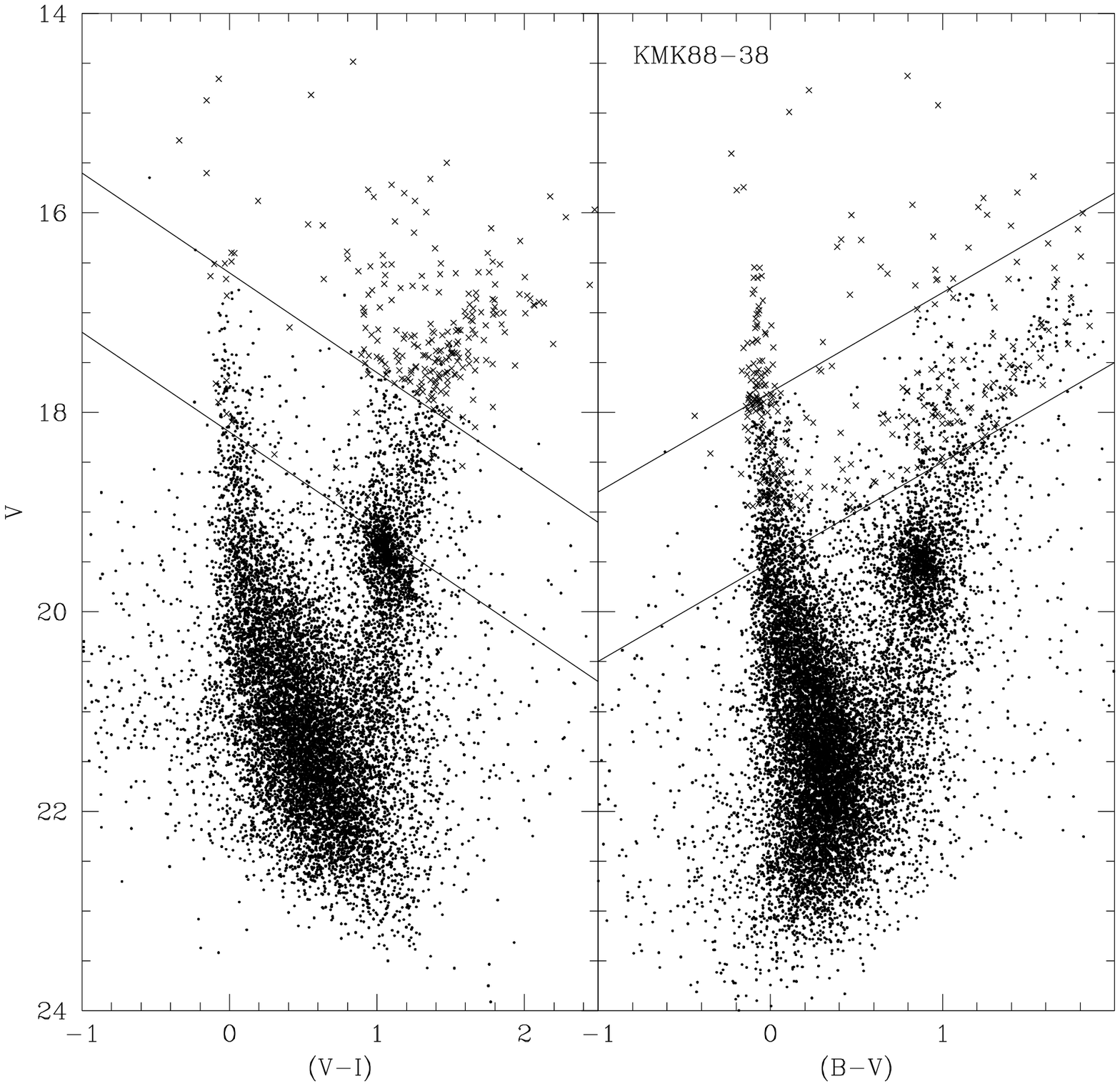}
 \includegraphics[width=0.4\textwidth]{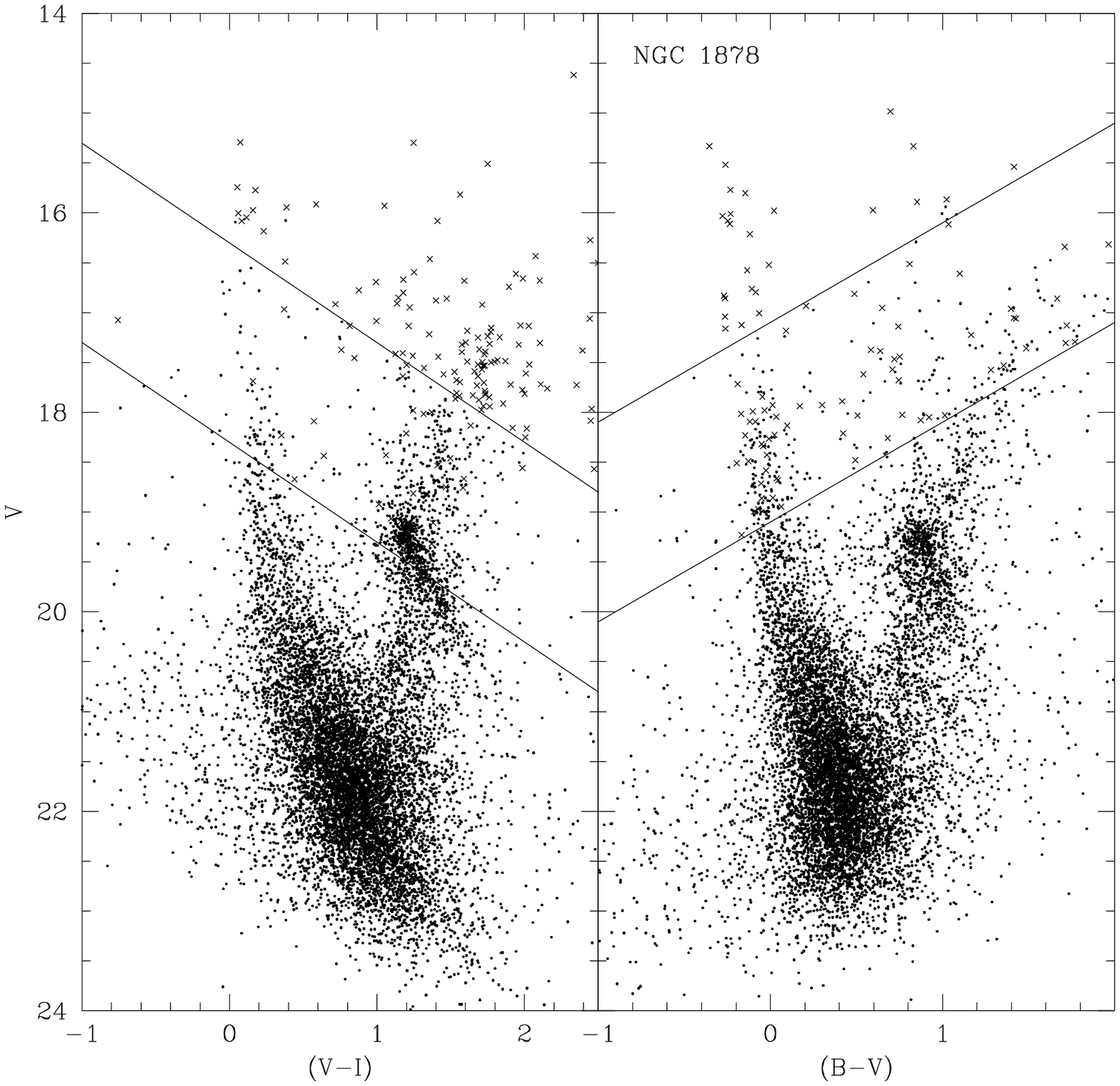}
 \caption{V,(B-V) (right panel) and V,(V-I) (left panel) CMDs from our sample. 
 The combined \textit{long} and \textit{short} CMD for the entire SOI field 
 is shown. The $\times$ are the stars that come from the \textit{short} 
 photometry, and the black dots are from the \textit{long} long photometry. 
 The upper solid line indicates the saturation limit and the lower line indicates the 
 faint magnitudes limit. The name of the cluster associated to the field is indicated 
 in each panel.}
 \label{cmdmerge}
\end{figure*}

In Fig. \ref{cmdmerge} we show the result of the merging process. 
We also display the \textit{long} saturation and \textit{short} faint 
magnitude limits computed as previously described (solid lines). The CMDs display a characteristic shape, being
dominated by the LMC fields stars, as expected, since our targets 
are intermediate to low-mass clusters. Moreover, we can distinguish the old and 
young population of the LMC divided by the extended Hertzsprung-gap, 
and we can identify a very clear Red Giant Branch (RGB) followed by the 
Red Clump (RC) and the Asymptotic Giant Branch (AGB). On the blue side of
the CMDs the Main-Sequence (MS) extend to the domain of 
bright magnitudes, showing the presence of younger populations.

\section{Analysis}

Since the clusters here studied are poorly populated objects, as shown in Fig. \ref{campos}, it is 
crucial to differentiate the star cluster population from the
field population. In fact, the field decontamination is the most
challenging step in our analysis. For instance, KMK88-38 appears on 
the image as an excess of only $\sim$ 13 \% relative to the same
area in the field. 

One popular method of decontamination is the 
statistical comparison of CMDs such as the one developed by 
\citet{kerber02}. In the literature other methods are
employed with success such as the statistical methods of \citet{gallart03,bb07,piatti08} and
the proper motion analysis made by \citet{richer}.

\subsection{Decontamination}

The statistical method of \citet{kerber02} is based on the fact that the
cluster population should appear as an over density in the 
colour-magnitude space (CMS) when compared against field stars. 
The decontamination is made by dividing the CMD into bins of colour
and magnitude and then comparing star counts in each bin in the cluster region 
with the corresponding bin in the field region. This method 
deals with very sharp boundaries when counting stars, which leads
to some shot noise when working with small numbers. The method uses a random process 
to select whether each star is a cluster member or not. This 
also leads to fluctuations on the final decontamination results.

An obvious improvement is to add an extra colour axis to the statistical field
subtraction process. In this work we explore two colours simultaneously and develop
a method of decontamination that operates on the colour-colour-magnitude 
space (CCMS). To avoid sharp boundaries in the CCMS bins, each star is replaced 
by a 3 dimensional Gaussian distribution with standard deviation equal to 
its photometric error. These additional features bring all the information 
available from the photometry into the analysis, using not only the position of 
the star in the CCMS but also its associated photometric error. 

The cluster region and center are determined by eye. 
Typical values for the cluster region radii are of $\sim 20\arcsec$ which corresponds 
to a fraction of $\sim 1/60$ of the SOAR/SOI FOV. In general the clusters 
are too sparse and low density to allow a more systematic approach. We use
all the remaining SOI area as the field region. For both the cluster and 
field regions we take the following steps:
(i) The CCMS is divided in a very fine grid of 0.01 in magnitude
and 0.005 in both colours.
(ii) For each bin we compute the contribution of each star by integrating its 3d-Gaussian 
over the bin volume. 
(iii) The resulting star counts in each bin are normalized by the different region areas. 

After applying step (iii) we subtract the field star counts from the cluster ones
at each bin. We then smooth out the resulting decontaminated cluster counts on CCMS 
by re-binning them on a coarser grid. For visualization purposes, we project the coarse CCMS 
grid onto the usual (V-I),V and (B-V),V CMD planes. The coarse grid size is chosen
as the smallest possible that is able to evidence a population that resembles a SSP with
minimal noise and acceptable resolution in the CMD plane. As we replaced a point process by smoothed 
3d-Gaussians, the resulting CMDs are expressed in terms of star counts per bin.

\subsection{Controlled experiments}

In order to test the developed decontamination algorithm we performed controlled 
experiments using artificially generated CMDs with parameters similar to
those expected for the clusters in our sample.

{To generate an artificial CMD, we used the aforementioned software 
developed by \citet{kerber02}. This algorithm takes into account 
realistic photometric errors and the effects of CMD broadening  
due to unresolved binaries; the adopted binary fraction adopted here is $50\%$.

We generated artificial CMDs for two isocrones models: one with 
$log(age) = 9.1$ and $Z = 0.010$ for which 50 stars were created; 
the other with $log(age) = 8.4$ and $Z = 0.010$ and containing 30 stars. 
15 realizations of each model were created.
Each artificial SSP realization is inserted into the real images 
at randomly chosen positions. By doing this we take into account 
effects caused by the stochastic nature of the artificial SSPs and the field. 

In Figure \ref{50rel} we show the resulting decontaminated V,(V-I) CMDs for the first 8 of 
the 15 controlled experiments from the first isocrone mentioned above 
($log(age) = 9.1$ and $Z = 0.010$, solid line). The artificial stars generated 
are shown as crosses in all panels. The mean contrast relative to
the field is $17\%$. The parameters for this model are typical 
for the expected oldest clusters 
in our sample and should reflect the quality of the decontamination 
algorithm for intermediate age SSPs. We show only 8 realizations of 
this experiment for simplicity.
 
Figure \ref{30rel} shows the results of the controlled experiment
for the other isochrone model, with a younger age and same metallicity.
The isochrone and artificial stars are again shown in all 8 panels.
The adopted age and metallicity ($log(age) = 8.4$ and $Z = 0.010$)
are consistent with the youngest clusters in our sample. To test
the method at yet lower constrast levels, only 30 stars where generated,
yielding a mean contrast of $11\%$.

The algorithm is successfull in recovering the enhanced 
density contrast in the CCMS regions associated to the input SSP.
Both the input Main Sequence Turn-Off (MSTO) and the RC positions can be 
identified in almost all realizations of the decontamination process 
applied to the first isochrone model (Figure \ref{50rel}). 
The extended MS locus associated to the young SSP is also clearly
visible in all but one of the panels of Figure \ref{30rel}. 
The dense contaminating field and its associated fluctuations, however,
often leave extra overdense CMD locci in some cases. In particular,
in a couple of the realizations shown, the CMD is dominated by a
faint and red plume of stars. This is likely caused by 
large and unaccounted for photometric errors at these faint magnitudes and 
red colours. Considering that this plume is not associated to any
reasonable SSP at the LMC distance, it can be easily discarded. More troubling
is the residual field SGB/RGB overdensity present in a couple of realizations 
shown in Figure \ref{30rel}. This could make the age estimate 
an ambiguous task. On the other hand, addition of an extra colour,
as is our case, helps remove this potential ambiguity. Also, in the vast 
majority of the realizations in both 
models it is possible to identify the input SSP as the dominant 
excess of stars in the CMDs,
to which isochrones may be visually overlaid in order to estimate their 
parameters.

\begin{figure}
 \includegraphics[angle=-90, width=0.48\textwidth]{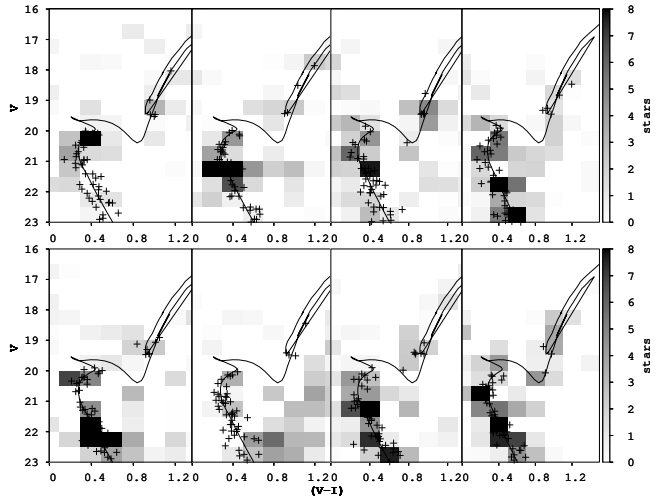}
 \vspace{12pt}
 \caption{$V,(V-I)$ CMDs resulting from controlled experiments 
using an intermediate age SSP with 50 stars. The employed isochrone is 
shown as the solid line in all panels. Its age is 1.25 Gyr and its
metallicity is half the solar value. The artificially generated stars 
are shown as crosses. The field subtracted star 
counts are coded with a grey scale, which is shown as a colour-bar on the 
right side. The bin sizes in color and magnitude are 
0.2 and 0.5 mag respectively. }
 \label{50rel}
\end{figure}

\begin{figure}
 \includegraphics[angle=-90, width=0.48\textwidth]{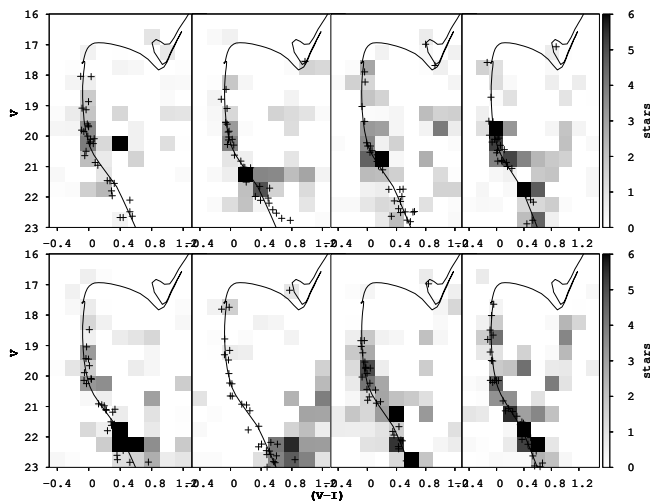}
 \vspace{12pt}
 \caption{$V,(V-I)$ CMDs resulting from controlled experiments using a
young SSP with 30 stars. The employed isochrone is shown in all panels as the
solid line. It has an age of 250 Myrs and a metallicity of half the solar 
value. The artificially generated stars are shown as the crosses. The bin 
size and grey scale representation of the field subtracted star 
counts are the same as in the previous figure. }
 \label{30rel}
\end{figure}

\subsection{Parameter determination}

CMDs are widely used as a tool for parameter determination 
in the literature \citep{kerber07,piatti07}, although there are
alternative methods that can be employed in order to obtain 
similar results. These methods include integrated spectroscopy
such as in \citet{ahumada,santos} and integrated photometry 
as in \citet{hunter,pessev}.

Figures \ref{field1} to \ref{field3} show the CMDs that result from the decontamination
process applied to our sample clusters. They can be used to extract physical parameters for the clusters. 
The target clusters are poorly populated, having typically from 20 to 150 stars. This limits
the accuracy of cluster parameters determination. Ideally, a
statistical method based on CMD star counts should be used 
to infer ages, metallicities, distances and redenning. 
This has been done by several of the authors quoted in this paper. 
However, in the current case, we showed in the previous subsection that 
the fluctuation in the background field is 
the dominant source of noise in the decontaminated CMDs. 
As we cannot single out and eliminate {\it a priori} 
the contaminating field stars present in the cluster direction, the use 
of statistical methods for CMD comparison
will not provide a significant advantage over isochrone fitting, as neither
is capable of eliminating the effect of field contamination on the defined
cluster region. This is a different situation from \citet{balbinot,kerber07}, 
for instance, who used statistical CMD comparisons to infer parameters 
and their uncertainties for rich star clusters. Therefore, a visual isochrone fit is adequate 
to estimate cluster parameters in the present case. Isochrones come from the stellar evolutionary 
models by \citet{leo}\footnote[2]{http://stev.oapd.inaf.it/cgi-bin/cmd}. 

We chose Padova isochrones although we are aware that systematic 
differences in the parameters may arise from
the choice of one model or another. This issue has been investigated in more
detail by \citet{kerberiau}, who compared different
stellar evolution models applied to LMC clusters. They show that, in relation 
to Padova isochrones, the ones from BaSTI \citep{basti} tend to overestimate the age 
by $\sim 0.3 Gyrs$. PEL \citep{pel} isochrones may lead
to a discrepancy in the distance of $\sim 2.8 kpc$ at $50 kpc$. The 
effects of convective overshooting (CO) are analysed in the same work. 
Obvious differences in age arise between CO models and non-CO models, 
in the sense that the former yield an older age when compared to the
latter ($log(age)_{(CO)} - log(age)_{(non-CO)} \simeq 0.1$). Apart from that,
however, \citet{kerberiau} see no other significant trend in the 
parameters. The authors also compared their metallicity values with those 
obtained from spectroscopic data \citep{olsz,groch} and found that the Padova models 
give the best agreement between spectroscopic and photometric data. We thus conclude that,
 for the purpose of this paper, the influence of the model adopted is negligible and 
 the differences from model to model are much smaller than the uncertainty of our parameter 
 determination.

For each cluster we derive the following parameters: age, metallicity, distance 
and colour excess. The parameters must obey some constraints in metallicity and 
distance, which are typical of LMC clusters. The allowed ranges adopted here are: 
$18.25 \leqslant (m-M)_0 \leqslant 18.75$ and 
$\frac{1}{3}Z_{\odot} \leqslant Z \leqslant Z_{\odot}$ \citep{kerber07}. We take the reddening 
tables from \citet{schlegel} as a initial guess of $E(B-V)$.
In some cases, the lack of clear CMD structures, such as the MSTO, RGB and RC prevent 
strong constraints on the parameters, specially the distance modulus. In such cases
we keep it fixed at the mean LMC value. 
 
The results of our parameter determination are listed in Table \ref{results}.
We will now comment the results for each cluster set. In Fig. \ref{field1} to 
\ref{field3} we show, along with the subtracted CMDs, three isochrones 
overploted (\textit{solid lines}). All isocrones have the same metallicity and 
only differ by age, whose values are given on the figures. On all figures the V,(B-V) CMD is on the left side
and the V,(V-I) is on the rigth side. The cluster names and other isochrone fit parameters
are also shown. All CMDs show star count in
bins of colour-magnitude instead of stars itself, the grey scale is adjusted 
to minimize the noise and increase the contrast for better visualization. 

\subsubsection{OGLE-LMC0531 and OGLE-LMC0523}

\begin{figure}
\centering
 \includegraphics[width=0.5\textwidth]{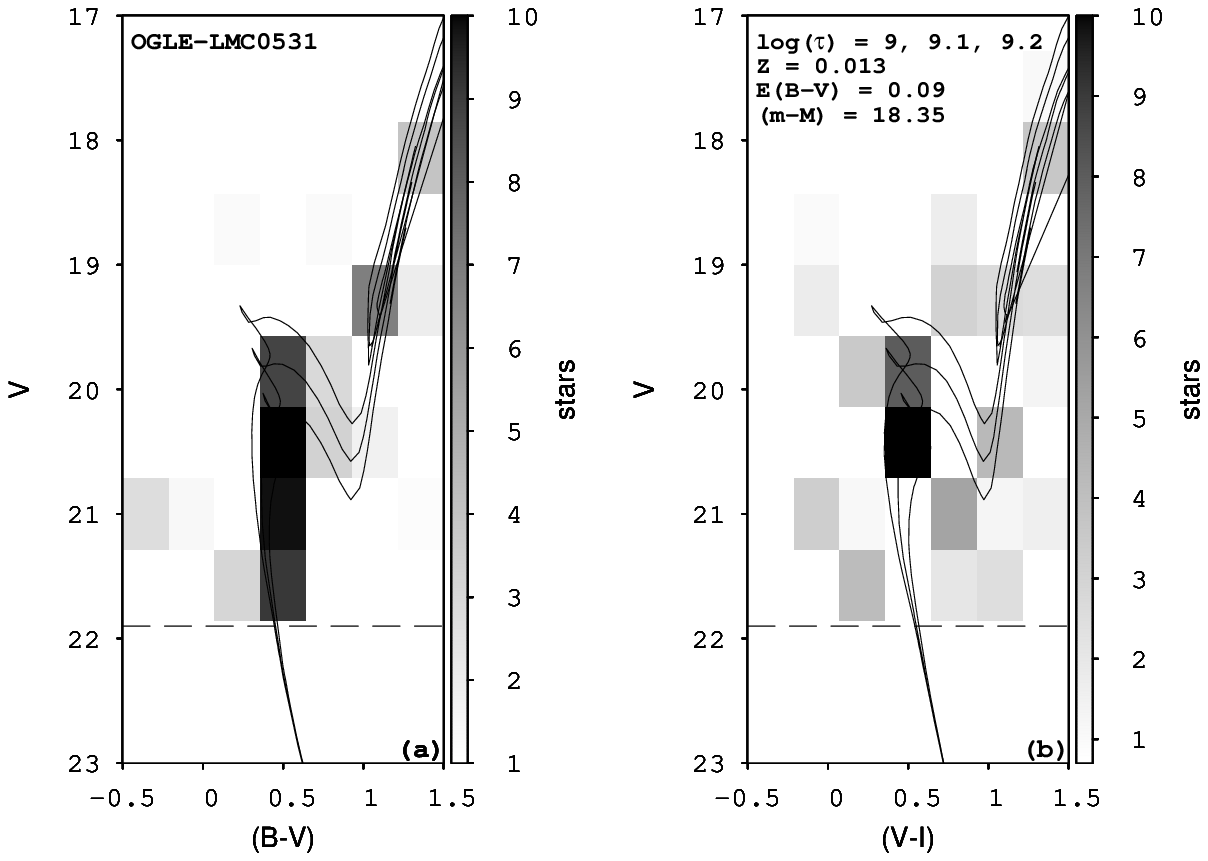}
 \includegraphics[width=0.5\textwidth]{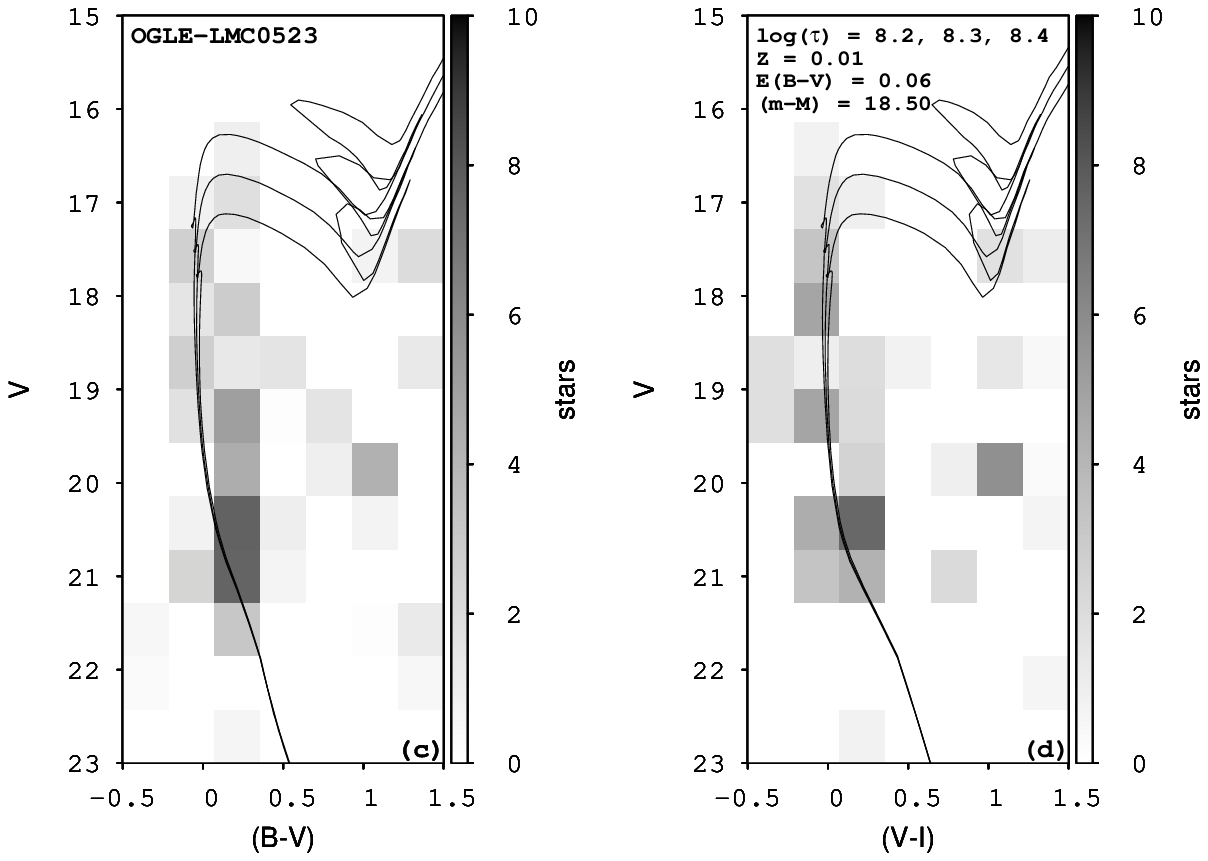}
 \caption{Panel {\it a}: $V,(B-V)$ CMD of OGLE-LMC0531; panel {\it b}:
 $V,(V-I)$ CMD for the same cluster as in previous panel. The same
 model, whose isochrone parameters are indicated, was used on both CMDs.
 Panel {\it c}: $V,(B-V)$ CMD of OGLE-LMC0523; panel {it d}: $V,(V-I)$
 CMD for the same cluster as in panel {\it c}. The same model, whose
 isochrone parameters are indicated, were used on both CMDs.
 In all panels, the field subtracted star counts are coded with a grey
 scale, which is shown as a colour-bar on the right side. The adopted
 bin size is 0.55 in magnitude and 0.30 in colour. The dashed line in the upper panels 
 marks the position of the MSTO of a cluster with $3Gyrs$ and half solar metallicity.}
 \label{field1}
\end{figure}

These two clusters are on the same SOAR/SOI field. Only the first one is
an age-gap candidate. As seen in Fig. \ref{field1}, the two clusters have
markedly distinct ages. In fact, OGLE-LMC0531 is older, with an age
between $\sim 1 - 2 Gyrs$. No sign of an upper MS is seen in either
the ${\it V,(B-V)}$ or ${\it V,(V-I)}$CMDs. In the former, its MSTO and
lower MS are more clearly defined, along with a trace of its RC. As
explained earlier, this allowed better constraints on its physical
parameters, yielding a $E(B-V) \simeq 0.09$, $(m-M)_{0} = 18.35$ and a
$Z = 0.013$. The extinction value is very similar to the initial
guess taken from \citet{schlegel}.

OGLE-LMC0523 has an age of $\sim 200 Myrs$. In reality, this is
an upper limit, since no clear turn-off is seen up to the
saturation limits. Since no other clear CMD structure besides the MS is
detected, it was harder to constrain the cluster distance.
We adopted the mean LMC distance of $(m-M)_{0} = 18.50$ and
found $Z = 0.010$ and $E(B-V) \sim 0.06$.

\subsubsection{KMK88-38, KMK88-39 and OGLE-LMC0214}

\begin{figure}
\centering
 \includegraphics[width=0.5\textwidth]{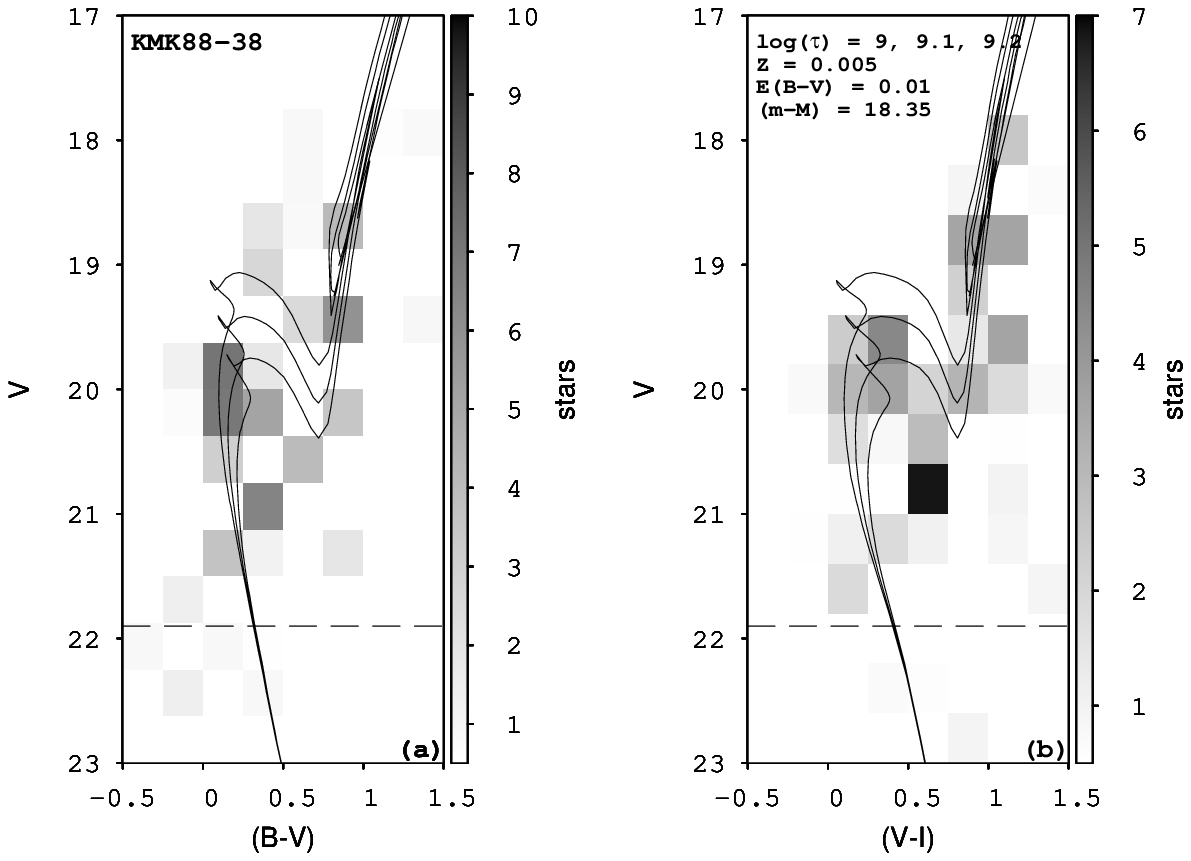}
 \includegraphics[width=0.5\textwidth]{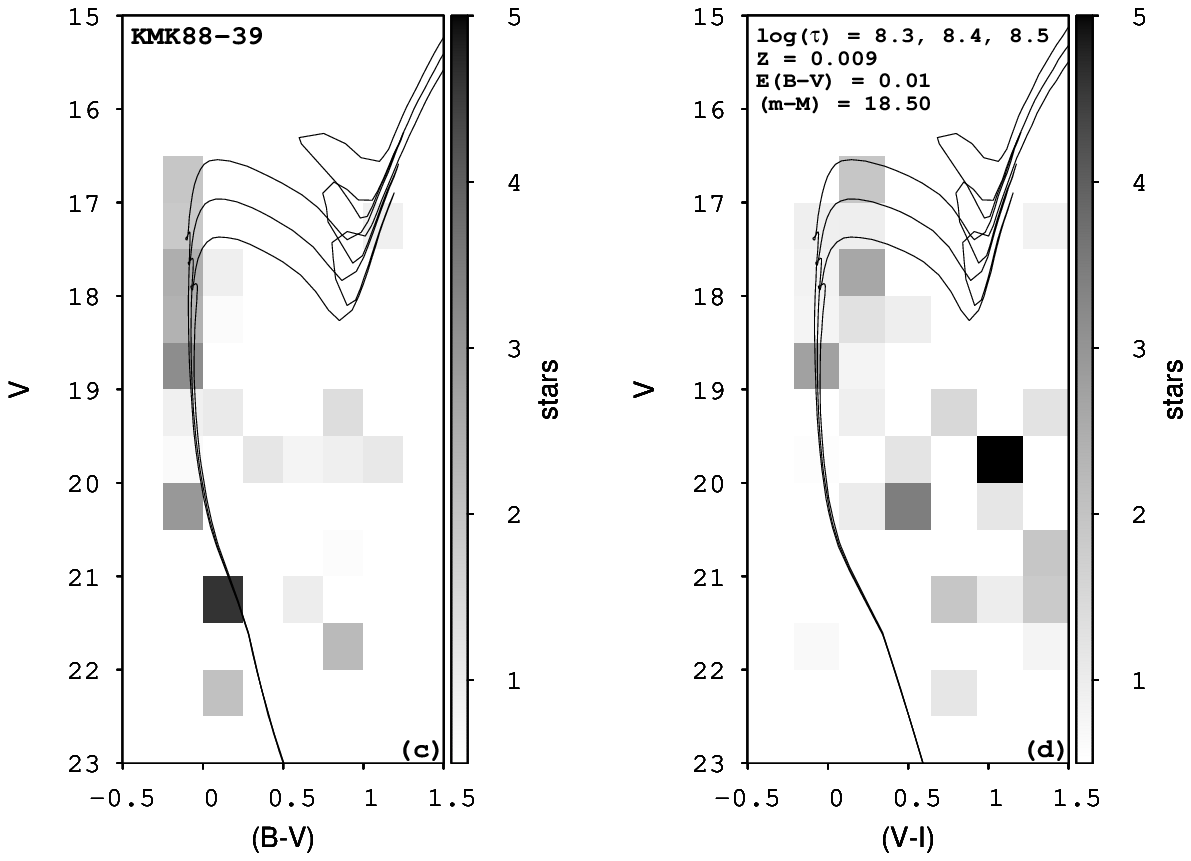}
 \caption{Same as in Fig. \ref{field1} but for the clusters
 KMK88-38 (panels {\it a} and {\it b}) and KMK88-39 (panels {\it c} and
{\it d}). The adopted bin size is 0.40 in magnitude and 0.25 in colour}
 \label{field2}
\end{figure}

KMK88-38 is our original target in this field, as an age-gap candidate.
The other clusters present on the field are KMK88-39 and OGLE-LMC0214.
This latter was imaged only in the V and I bands, due to variations
in telescope pointing. Even though we do not show its CMDs, it has been
submitted to the same analysis as the other clusters, the results of
which are shown in Table \ref{results}.

The field subtracted CMDs for KMK88-38 and KMK88-39 are shown
in Fig. \ref{field2}. The panels and symbol conventions
are the same as in Fig. \ref{field1}. KMK88-38 is relatively old,
with an age between $1 \sim 2 Gyrs$, similar to OGLE-LMC0531.
The presence of strong CMD features, such as a MSTO, and an RGB allowed
us to explore a wider range of parameters with the effect of better
constraining their determination. The derived parameters for KMK88-38
are: $(m-M)_{0} = 18.35$, $Z = 0.005$ and $E(B-V) = 0.02$.

On the other hand, KMK88-39 is very young, with an age of $\sim 250
Myrs$ when using a metallicity of $Z = 0.009$. The age derived for
this cluster is again an upper limit, since no clear MSTO is seen.
The distance adopted was again the mean LMC distance, and we found
a redenning of $E(B-V) = 0.01$.

\subsubsection{NGC 1878}

NGC 1878 was accidentally observed in place of the age gap candidate
cluster BSDL917. In contrast to the other fields,
only one cluster was imaged this time. Even though is an NGC
object, NGC 1878 had never been studied before.

Since NGC 1878 is a relatively rich cluster ($\gtrsim 100$ stars)
it allowed us to explore our decontamination algorithm in a
simpler case, where the density of cluster stars in the CMD at a given
colour-magnitude bin is significantly greater than the noise generated
by the fluctuation of field stars in this same bin. In Fig. \ref{field3}
we show its subtracted CMD, adopting the same panels and symbols
as in the two previous figures. We clearly
see much more well defined CMD loci than in the other cases,
indicating that the decontamination process is effective.

The {\it V,(B-V)} CMD of NGC 1878 looks slightly bluer than expected
from the {\it V,(V-I)} CMD. This means that a unique isochrone solution
to both CMDs was harder to achieve. As a compromise, the best visual
fit solution is slightly red (blue) when compared to the main MS locus
in the former (latter) CMD. This effect can result from a combination
of crowding and variable PSF, which may lead to increased light
contamination on PSF fitting to the B band, therefore making the
star seem bluer. The best isochrone fit gives a young age, $\sim
200 Myrs$, a typical LMC metallicity of
$Z = 0.009$, $E(B-V) = 0.12$. We adopted a distance modulus
of $(m-M)_{0} = 18.50$.

\begin{figure}
\centering
 \includegraphics[width=0.5\textwidth]{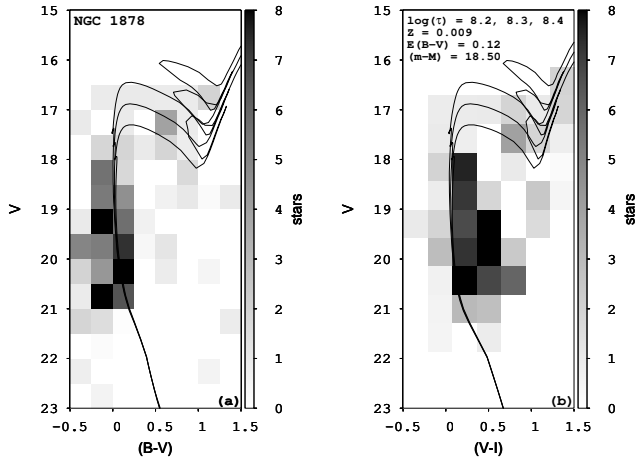}
 \caption{Same as in Fig. \ref{field1} but for the cluster
 NGC 1878. The adopted bin size is 0.50 in magnitude and 0.25 in colour}
 \label{field3}
\end{figure}

\begin{table}
\centering
\begin{tabular}{lcccc}

\hline
$ Name $&$ log(Age) $&$ Z $&$ E(B-V) $& $(m-M)_0 $\\
\hline
\hline
OGLE-LMC0214 &$ 8.4 \downarrow $&$ 0.013 $&$ 0.10 $ &$ 18.50* $\\ 
OGLE-LMC0523 &$ 8.3 \downarrow $&$ 0.010 $&$ 0.06 $ &$ 18.50* $\\ 
OGLE-LMC0531 &$ 9.1 \pm 0.1    $&$ 0.013 $&$ 0.09 $ &$ 18.35 $\\ 
KMK88-38     &$ 9.1 \pm 0.1    $&$ 0.005 $&$ 0.02 $ &$ 18.35 $\\ 
KMK88-39     &$ 8.4 \downarrow $&$ 0.009 $&$ 0.01 $ &$ 18.50* $\\ 
NGC1878      &$ 8.3 \pm 0.2    $&$ 0.009 $&$ 0.12 $ &$ 18.50* $\\ 
\hline 
\end{tabular} 
\caption{The result of our CMD analysis. We list the object name, the 
derived log(age), metalicity, $E(B-V)$ and the distance modulus. The 
arrows indicate upper limits. The asterisk indicates
where the distance modulus was fixed at the given value.}
\label{results}
\end{table}

\section{Conclusions}

We have obtained SOAR/SOI images of 3 fields containing 6 LMC clusters,
two of which were previously identified as candidates to fill the LMC age-gap. 
Photometry was carried out in three filters (B,V and I) reaching well below 
the MSTO of an old stellar population. 

All but one of these clusters are poor and sparse, consistent with
intermediate to low initial masses, requiring a careful and objective
process to eliminate field stars from the cluster CMD. We developed such method
that simultaneously uses the information in the available three-dimensional 
space of magnitudes and colours. With the resulting decontaminated CMDs we
carried out visual isochrone fits based on Padova evolutionary sequences. 

Our main result is that none of the clusters here studied is older than
2 Gyrs, therefore not filling the LMC age-gap. The two 
main candidates, OGLE-LMC0531 and KMK88-38, are in fact the oldest 
in the sample,
displaying a clear MSTO and a branch of evolved stars. Their
MSTO magnitudes are about 2 mag brighter than that expected
for a 3 Gyrs old SSP at the LMC distance, as evidenced from 
Figures \ref{field1} and \ref{field2}. Interestingly, they 
are sided on the images by much younger clusters, for which only upper 
limits in age ($< 200 Myrs$) could be derived from the CMDs. This large 
age range is evidence either of a complex cluster formation history, or
very efficient orbital mixing within the LMC. Another noticeable result
is that the ages here inferred are similar to the last two close encounters
between the LMC and SMC according to recent N-body 
simulations \citep{gardiner}, adding a few more examples of 
clusters that likely have formed as the result of SMC/LMC close-by passages.
We also note that the two older clusters are systematically on the foreground 
relative to current models for the LMC disk \citep{kerber07,nikolaev}.

The clusters in our sample are of lower initial mass 
($M \sim 10^4 M_{\odot}$) than most LMC clusters studied so far.
The fact that they do not fill the age gap may indicate that
yet lower mass clusters, possibly cluster remnants, must be sampled and 
studied in order to bridge the apparent inconsistency between 
reconstructed cluster and field star formation histories in the LMC.

{\bf Acknowlegments.}This work was supported by Conselho Nacional de Desenvolvimento 
Cient\'{i}fico e Tecnol\'{o}gico (CNPq) in Brazil and Coordena\c{c}\~{a}o 
de Aperfei\c{c}oamento de Pessoal de N\'{i}vel Superior.

\label{lastpage}


\begin{thebibliography}{}
\bibitem[\protect\citeauthoryear{Ahumada et 
al.}{2009}]{ahumada} Ahumada A.~V., Talavera M.~L., Clari{\'a} 
J.~J., Santos J.~F.~C., Bica E., Parisi M.~C., Torres M.~C., 
2009, IAUS, 256, 293 
\bibitem[\protect\citeauthoryear{Balbinot et al.}{2009}]{balbinot} 
Balbinot E., Santiago B.~X., Bica E., Bonatto C., 
2009, MNRAS, 396, 1596 
\bibitem[\protect\citeauthoryear{Baume et al.}{2008}]{baume} 
Baume G., No{\"e}l N.~E.~D., Costa E., Carraro G., M{\'e}ndez R.~A., Pedreros M.~H., 
2008, MNRAS, 390, 1683 
\bibitem[\protect\citeauthoryear{Bica et al.}{1999}]{bica99} 
Bica E.~L.~D., Schmitt H.~R., Dutra C.~M., Oliveira H.~L., 
1999, AJ, 117, 238 
\bibitem[\protect\citeauthoryear{Bonatto \& Bica}{2007}]{bb07} 
Bonatto C., Bica E., 
2007, MNRAS, 377, 1301 
\bibitem[\protect\citeauthoryear{Castellani et al.}{2003}]{pel} 
Castellani V., Degl'Innocenti S., Marconi M., Prada Moroni P.~G., Sestito P., 2003, A\&A, 404, 645 
\bibitem[\protect\citeauthoryear{de Grijs \& Anders}{2006}]{grijs} 
de Grijs R., Anders P., 
2006, MNRAS, 366, 295 
\bibitem[\protect\citeauthoryear{Gallart et al.}{2003}]{gallart03}
Gallart C., Zoccali M., Bertelli G., Chiosi C., Demarque P., Girardi L.,
Nasi E., Woo J.-H., Yi S., 2003, AJ, 125, 742
\bibitem[\protect\citeauthoryear{Gardiner, Sawa, \& Fujimoto}{1994}]{gardiner} 
Gardiner L.~T., Sawa T., Fujimoto M., 1994, MNRAS, 266, 567
\bibitem[\protect\citeauthoryear{Girardi et al.}{2002}]{leo} 
Girardi L., Bertelli G., Bressan A., Chiosi C., Groenewegen M.~A.~T., Marigo P., Salasnich B., Weiss A., 
2002, A\&A, 391, 195 
\bibitem[\protect\citeauthoryear{Grocholski et al.}{2006}]{groch} 
Grocholski A.~J., Cole A.~A., Sarajedini A., Geisler D., Smith V.~V. 2006, AJ, 132, 1630
\bibitem[\protect\citeauthoryear{Holtzman et al.}{1999}]{holtzman}
Holtzman J.~A., et al., 
1999, AJ, 118, 2262 
\bibitem[\protect\citeauthoryear{Hunter et al.}{2003}]{hunter}
Hunter D.~A., Elmegreen B.~G., Dupuy T.~J., Mortonson M., 
2003, AJ, 126, 1836 
\bibitem[\protect\citeauthoryear{Javiel, Santiago, \& Kerber}{2005}]{javiel} 
Javiel S.~C., Santiago B.~X., Kerber L.~O., 
2005, A\&A, 431, 73 
\bibitem[\protect\citeauthoryear{Jensen, Mould, \& Reid}{1988}]{jensen} 
Jensen J., Mould J., Reid N., 
1988, ApJS, 67, 77 
\bibitem[\protect\citeauthoryear{Kerber \& Santiago}{2009}]{kerberiau} 
Kerber L.~O., Santiago B.~X., 2009, IAUS, 256, 391 
\bibitem[\protect\citeauthoryear{Kerber, Santiago, \& Brocato}{2007}]{kerber07}
Kerber L.~O., Santiago B.~X., Brocato E., 
2007, A\&A, 462, 139 
\bibitem[\protect\citeauthoryear{Kerber et al.}{2002}]{kerber02}
Kerber L.~O., Santiago B.~X., Castro R., Valls-Gabaud D., 
2002, A\&A, 390, 121 
\bibitem[\protect\citeauthoryear{Kontizas, Metaxa, \& Kontizas}{1988}]{kmk}
Kontizas E., Metaxa M., Kontizas M., 
1988, AJ, 96, 1625
\bibitem[\protect\citeauthoryear{Nikolaev et al.}{2004}]{nikolaev}
Nikolaev S., Drake A.~J., Keller S.~C., Cook K.~H., Dalal N., Griest K., Welch D.~L., Kanbur S.~M., 
2004, ApJ, 601, 260 
\bibitem[\protect\citeauthoryear{No{\"e}l et al.}{2007}]{noel}
No{\"e}l N.~E.~D., Gallart C., Costa E., M{\'e}ndez R.~A., 
2007, AJ, 133, 2037 
\bibitem[\protect\citeauthoryear{Olszewski et al.}{1991}]{olsz}
Olszewski E.~W., Schommer R.~A., Suntzeff B., Harris H. 1991, AJ, 101, 515
\bibitem[\protect\citeauthoryear{Parmentier \& de Grijs}{2008}]{parm} 
Parmentier G., de Grijs R., 
2008, MNRAS, 383, 1103 
\bibitem[\protect\citeauthoryear{Pessev et al.}{2008}]{pessev} 
Pessev P., Goudfrooij P., Puzia T., Chandar R., 
2008, AAS, 211, \#162.27 
\bibitem[\protect\citeauthoryear{Piatti et al.}{2009}]{piatti09}
Piatti A.~E., Geisler D., Sarajedini A., Gallart C., 
2009, A\&A, 501, 585 
\bibitem[\protect\citeauthoryear{Piatti et al.}{2008}]{piatti08}
Piatti A.~E., Geisler D., Sarajedini A., Gallart C., Wischnjewsky M., 
2008, MNRAS, 389, 429
\bibitem[\protect\citeauthoryear{Piatti et al.}{2007}]{piatti07} 
Piatti A.~E., Sarajedini A., Geisler D., Gallart C., Wischnjewsky M., 
2007, MNRAS, 382, 1203 
\bibitem[\protect\citeauthoryear{Pietrinferni et al.}{2004}]{basti} 
Pietrinferni A., Cassisi S., Salaris M., Castelli F., 2004, ApJ, 612, 168 
\bibitem[\protect\citeauthoryear{Pietrzynski et al.}{1999}]{ogle}
Pietrzynski G., Udalski A., Kubiak M., Szymanski M., Wozniak P., Zebrun K., 
1999, AcA, 49, 521 
\bibitem[\protect\citeauthoryear{Rich, Shara, \& Zurek}{2001}]{rich}
Rich R.~M., Shara M.~M., Zurek D., 
2001, AJ, 122, 842 
\bibitem[\protect\citeauthoryear{Richer et al.}{2008}]{richer} 
Richer H.~B., et al., 
2008, AJ, 135, 2141 
\bibitem[\protect\citeauthoryear{Santiago}{2009}]{santiagoiau} 
Santiago B.~X., 2009, IAUS, 256, 69 
\bibitem[\protect\citeauthoryear{Santos et al.}{2006}]{santos} 
Santos J.~F.~C., Jr., Clari{\'a} J.~J., Ahumada A.~V., Bica E., Piatti A.~E., Parisi M.~C., 
2006, A\&A, 448, 1023 
\bibitem[\protect\citeauthoryear{Sarajedini}{1998}]{saraj} 
Sarajedini A., 
1998, AJ, 116, 738 
\bibitem[\protect\citeauthoryear{Schlegel, Finkbeiner, \& Davis}{1998}]{schlegel}
Schlegel D.~J., Finkbeiner D.~P., Davis M., 
1998, ApJ, 500, 525 
\bibitem[\protect\citeauthoryear{Sharpee et al.}{2002}]{sharpee}
Sharpee B., Stark M., Pritzl B., Smith H., Silbermann N., Wilhelm R., Walker A., 
2002, AJ, 123, 3216 
\bibitem[\protect\citeauthoryear{Stetson}{1994}]{stetson} 
Stetson P.~B., 1994, PASP, 106, 250 
\end{thebibliography}
\end{document}